\documentclass[preprint,11pt]{aastex}

\slugcomment{Accepted for publication in The Astrophysical
Journal Supplement Series}
\shorttitle{{\it FUSE} Translucent H$_2$ Survey}
\shortauthors{Rachford et al.}

\begin{document}

\title{Molecular Hydrogen in the {\it FUSE} Translucent Lines of Sight: The Full Sample}

\author{Brian L. Rachford\altaffilmark{1},
Theodore P. Snow\altaffilmark{2},
Joshua D. Destree\altaffilmark{2},
Teresa L. Ross\altaffilmark{2},
Roger Ferlet\altaffilmark{3},
Scott D. Friedman\altaffilmark{4},
Cecile Gry\altaffilmark{5},
Edward B. Jenkins\altaffilmark{6},
Donald C. Morton\altaffilmark{7},
Blair D. Savage\altaffilmark{8},
J. Michael Shull\altaffilmark{2},
Paule Sonnentrucker\altaffilmark{9},
Jason Tumlinson\altaffilmark{10},
Alfred Vidal-Madjar\altaffilmark{3},
Daniel E. Welty\altaffilmark{11},
Donald G. York\altaffilmark{11,12}
}

\altaffiltext{1}{Department of Physics, Embry-Riddle Aeronautical University, 3700 Willow Creek Road, Prescott, AZ 86301; rachf7ac@erau.edu}
\altaffiltext{2}{Center for Astrophysics and Space Astronomy, Department of Astrophysical and Planetary Sciences, University of Colorado, Boulder, CO 80309-0389}
\altaffiltext{3}{Institut d'Astrophysique de Paris, UMR7095 CNRS, Universite Pierre \& Marie Curie, 98bis Boulevard Arago, 75014 Paris, France}
\altaffiltext{4}{Space Telescope Science Institute, 3700 San Martin Drive, Baltimore, MD 21218}
\altaffiltext{5}{Laboratoire d'Astrophysique de Marseille, CNRS, 38 av. Frederic Joliot-Curie, 13388 Marseille cedex 13, France}
\altaffiltext{6}{Princeton University Observatory, Peyton Hall, Princeton, NJ 08544}
\altaffiltext{7}{Herzberg Institute of Astrophysics, National Research Council, 5071 W. Saanich Road, Victoria, BC V9E 2E7, Canada}
\altaffiltext{8}{Department of Astronomy, University of Wisconsin, 475 North Charter Street, Madison, WI 53706}
\altaffiltext{9}{Department of Physics and Astronomy, Johns Hopkins University, 3400 North Charles Street, Baltimore, MD 21218}
\altaffiltext{10}{Yale Center for Astronomy and Astrophysics, Department of Physics, P.O. Box 208121, New Haven, CT 06520}
\altaffiltext{11}{Department of Astronomy and Astrophysics, University of Chicago, 5640 South Ellis Avenue, Chicago, IL 60637}
\altaffiltext{12}{Enrico Fermi Institute, University of Chicago, 5640 South Ellis Avenue, Chicago, IL 60637}

\begin{abstract}
We report total abundances and related parameters for the full
sample of the {\it FUSE} survey of molecular hydrogen in 38
translucent lines of sight.  New results are presented for the
``second half'' of the survey involving 15 lines of sight to
supplement data for the first 23 lines of sight already
published.  We assess the correlations between molecular
hydrogen and various extinction parameters in the full sample,
which covers a broader range of conditions than the initial
sample.  In particular, we are now able to confirm that many,
but not all, lines of sight with shallow far-UV extinction
curves and large values of the total-to-selective extinction
ratio, $R_V$ = $A_V$/$E(B-V)$ --- characteristic of larger
than average dust grains --- are associated with particularly low
hydrogen molecular fractions ($f_{\rm H2}$).  In the lines of
sight with large $R_V$, there is in fact a wide range in
molecular fractions, despite the expectation that the larger
grains should lead to less H$_2$ formation.  However, we see
specific evidence that the molecular fractions in this
sub-sample are inversely related to the estimated strength of
the UV radiation field and thus the latter factor is more important
in this regime.  We have provided an update to previous values
of the gas-to-dust ratio, $N$(H$_{\rm tot}$)/$E(B-V)$, based on
direct measurements of $N$(H$_2$) and $N$(H I).  Although our
value is nearly identical to that found with {\it Copernicus}
data, it extends the relationship by a factor of 2 in reddening.
Finally, as the new lines of sight generally show low to
moderate molecular fractions, we still find little evidence for
single monolithic ``translucent clouds'' with $f_{\rm H2}$
$\sim$ 1.

\end{abstract}


\section{Introduction}
Molecular hydrogen is the most abundant molecule in the
universe, and a detailed knowledge of H$_2$ is crucial for a
full understanding of the physics and chemistry of the
interstellar medium.  A broad overview of interstellar H$_2$ is
provided by Shull \& Beckwith (1982), while a recent overview of
chemical processes in diffuse and translucent clouds is provided
by Snow \& McCall (2006).

A major goal of the {\it Far Ultraviolet Spectroscopic Explorer}
({\it FUSE}) was a survey of molecular hydrogen in 45 lines of
sight with an emphasis on interstellar clouds with as much
extinction as possible.  The extinctions covered the range of
the so-called ``translucent clouds'' with $A_V$ in the range
1--5 mag.  Within these limits, lines of sight were chosen from
a variety of environments and dust characteristics.  The first
results from the study were presented by Rachford et al.\ (2002
[Paper I]).

Paper I contains the details regarding the previous history of
H$_2$ observations and the rationale for the {\it FUSE} survey.
Slightly more than one-half of the planned targets had been
observed and analyzed and those lines of sight were detailed.  A
main finding was that in most cases a line of sight was likely
composed of multiple clouds, suggesting a change in terminology
to ``translucent lines of sight'' pending the clear
identification of a high-extinction line of sight made up of one
highly molecular cloud.

The main purpose of the present paper is to provide the
community with the overall H$_2$ results for the full sample, as
well as refine and strengthen some of the conclusions of Paper
I.  The additional lines of sight give us much better coverage
of the variety of dust characteristics than the original partial
sample and we emphasize those results.  Also, we have taken
advantage of the publication of the 2MASS All-Sky Survey of
Point Sources (Skrustkie et al.\ 2006) to provide more reliable
extinction parameters for the full sample.

The rest of the paper is organized as follows: in \S\ 2 we
describe the remaining target for the FUSE translucent H$_2$
survey, along with comments on the stars chosen; in \S\ 3 we
give the details of the observations, data reduction, and
analysis; in \S\ 4 we discuss the results and their
implications; and we summarize the paper in \S\ 5.

\section{Target Selection and Stellar Properties}

\subsection{General Comments on the Sample}
Specific information on the selection criteria is given in Paper
I, which includes details on the {\it FUSE} programs containing
the observed stars.  In this paper, we report new H$_2$ results
for observations towards 15 stars, whose basic parameters are
listed in Table 1.  In a few cases, the data quality was poor
for the new targets, but we were able to obtain reasonable H$_2$
measurements.

While the overall observing program was very successful, for 7
out of the original 45 targets {\it FUSE} was unable to
adequately observe the target, or not observe it at all, as
follows.  HD 37021 and HD 37061 are both very bright targets in
the Orion molecular cloud and the lines of sight display unusual
extinction characteristic of larger than normal dust grains.  We
had anticipated that the sensitivity of {\it FUSE} would
decrease to the point where the UV fluxes would be safe, but
that did not happen.  HD 147889 is in the $\rho$ Oph complex,
which also shows unusual extinction in the same sense as the
Orion targets.  However, a dearth of suitable guide stars in the
field prevented this observation from taking place.
Fortunately, {\it FUSE} was able to observe nearby HD 147888 as
noted below.  Walker 67 (in open cluster NGC 2264) and HD 166734
are two targets on the high end of the extinction range we
wished to cover, but both targets were fainter in the far UV
than anticipated and {\it FUSE} did not collect enough counts in
either observation for an adequate analysis.  HD 94414 was also
a faint target and had some data quality issues.  HD 21483 was
also not observed.

\subsection{Stellar and Extinction Properties}
We have generally applied the techniques of Paper I to determine
the relevant extinction parameters for each line of sight,
namely the color excess ($E(B-V)$), the total-to-selective
extinction ratio ($R_V$), and the total visual extinction
($A_V$).  However, in the interim between Paper I and the
current paper we have made some improvements.  Many of the
authors of both papers have been involved in a major project to
understand the diffuse interstellar bands (DIBs).  Part of this
project includes deriving a consistent set of extinction
measures for the {\it FUSE} translucent lines of sight plus a
much larger sample of reddened stars.  This project has led to
some modifications in the tabulated $V$ magnitudes, spectral
types, and extinction values used for our {\it FUSE} sample.
These modifications result in the best consistency between the
translucent sample, our DIB project, and the papers on chemical
depletions in the translucent sample in which the two lead
authors of the present paper were involved (Jensen, et al.\
2005, 2007).  We have thus not only provided extinction
parameters for the new targets, but also the revised values for
the targets from Paper I, and these values are given in Table 2.

The differences between the new procedure for determining
extinction parameters and the original procedures in \S\ 2.2 of
Paper I are as follows.  First, the values of $E(B-V)$ were
homogenized as much as possible, which has led mostly to
cosmetic changes from our values in Paper I.  Two exceptions are
HD 24534 and HD 110432, which are both emission-line stars where
some of the observed reddening is likely circumstellar, and the
$E(B-V)$ values we quote here are somewhat larger than those
given in Paper I.  However, these changes do not result in any
significant changes in the overall results.

Our primary technique for determining $R_V$ involves fitting a
functional form to color excess ratios at optical and IR
wavelengths.  This is based on the method of Martin \& Whittet
(1990), which we discuss in detail in Paper I.  Subsequent to
the completion of Paper I, the full version of the 2MASS All-Sky
Survey of Point Sources (Skrutskie et al.\ 2006) became
available which includes very high quality $JHK$ photometry for
all of our sources.  Thus, we have used these values in our
derivation of $R_V$, which has significantly improved the
quality, consistency, and completeness of extinction values in
our sample.  We also used correlations between $R_V$ and the
wavelength of maximum polarization ($R_V$ = 5.6$\lambda_{\rm
max}$, with $\lambda$ in $\mu$m; Whittet \& van Breda 1978) and
the far-UV rise in the extinction curve ($c_2$ = $-$0.824 +
4.717$R_V^{-1}$; Fitzpatrick 1999) as ancillary proxies for the
photometric values, again as discussed in Paper I.

The listed uncertainties in the extinction parameters in Table 2
are necessarily estimates, but are mostly based on the scatter
observed for that particular technique or correlation and the
notion that systematic effects are likely important.  For the
$R_V$ values that did not come from an analysis of the IR
photometry, and for the latter values when the fit appeared
qualitatively reasonable, we assumed uncertainties of 0.3.  In
several cases, the IR photometry deviated from the expected
relationship and we thus adopted larger error bars.  However, in
those cases, if at least one of the other two techniques for
determining $R_V$ agreed with the IR method, we then adopted the
IR value and used smaller error bars.  Finally, when we did not
have any $R_V$ information (only for HD 186994), we adopted the
Galactic average value of 3.1 (Draine 2003).  As can be seen in
Table 2, in only 3 cases, HD 24534, HD 102065, and HD 186994 did
we not use the $R_V$ derived from the IR photometry.  HD 24534
is a Be star (as discussed below) for which we could not confirm
the photometric value with other methods.  The other two stars
have $E(B-V)$ $<$ 0.2 for which we were concerned about the large
relative effect that a small error in the color indices or
spectral type will have on the photometric $R_V$.  Published
values of $R_V$ for HD 102065 are near 4.0 (Boulanger, et al.\
1994; Paper I).

It is worth noting that as we pointed out in Paper I, the
photometric method is the most direct and the other two methods
are basically correlations between another parameter and $R_V$
derived from IR photometry.  Thus, it is not a surprise that
there is generally excellent agreement between the three
indicators.  Of the 22 points with both photometric and
polarimetric measurements only 4 disagree by more than 2
$\sigma$, and only an additional 3 disagree by more than 1
$\sigma$.  The statistics comparing the photometric and
extinction curve values for $R_V$ are nearly identical.  One
limitation with the polarimetric method is that it is poorly
constrained for $R_V$ $\gtrsim$ 4 (Whittet \& van Breda 1978;
Figure 1).  In fact, we do not derive $R_V$ $>$ 3.9 for any of
our targets with this method, even when the other two methods
give larger $R_V$.  This is one reason we have not attempted to
average together values from different methods.

There are two particular problems with the photometric method of
determining $R_V$ that we wish to address.  First, 7 of the
stars in our sample are Be stars\footnote{In addition to the 5
Be stars in Table 1, HD 24534 and HD 110432 from Paper I are
also Be stars.} and thus have circumstellar material.  Processes
in this material skew not only the optical photometry via
emission lines and continuous free-bound emission, but also
optical polarization and IR photometry.  The other issue is
variability.  Not only are Be stars generally variable, but
there are other types of variable stars within the O and early B
types.  

In a similar context to ours, Bowen et al. (2008) discussed
these issues as they related to determining extinction
parameters for the {\it FUSE} survey of O VI in the Galactic
disk.  These targets generally had $E(B-V)$ $\lesssim$ 0.4.  One
advantage we have in the present survey is that our targets
generally have $E(B-V)$ $\gtrsim$ 0.4.  This is important
because the greater the color excess, the smaller the relative
error caused by variability, and variability is often small in
this stellar temperature range.  The Be stars are somewhat more
problematic, although again we expect the fractional effects in
our extinction parameters will be lessened for the more highly
reddened stars, and most of the Be stars have small variability
amplitudes (e.g. Hubert \& Floquet 1998).  Furthermore, issues
like these are one reason that we have typically adopted
relatively large errors for $R_V$.  In fact, all 7 of the Be
stars in our sample were flagged as not fitting the Martin \&
Whittet (1990) relation very well and thus were given
particularly large errors, which were only reduced in the
adopted $R_V$ if either of the other methods agreed with the
photometric value.  Still, because these values may be
unreliable for Be stars, in some cases we have eliminated these
stars when considering a particular correlation.

While preparing our present manuscript, Fitzpatrick \& Massa
(2005, 2007) published an updated version of their seminal work
in parametrizing extinction curves.  They slightly modified
their parametrization of the UV portion of the extinction
curves, and greatly expanded the sample of lines of sight,
including a few from our sample which have not previously been
analyzed.  However, since other authors have published curves in
the original 6-parameter scheme (Fitzpatrick \& Massa 1986,
1988, 1990), an exclusive use of the new scheme would result in
a significant decrease in the fraction of stars available in a
self-consistent system.  Thus, we have continued to use values
based on the older parametrization which are given in Table 3
for our new targets.  The key parameter for the present paper is
$c_2$, the linear coefficient of the far-UV rise in the curve.

We note that Fitzpatrick \& Massa (2005, 2007) also used 2MASS
$JHK$ photometry for the IR portion of the extinction curves,
and derived $R_V$ for their sample using a technique similar to
ours (Paper I).  Indeed, there is an excellent match between our
values and theirs, as expected.

\subsection{Special line-of-sight characteristics}
Many of the lines of sight are of special interest due to their
location and/or environment.  As in Paper I, we give a brief
overview of each line of sight in the following sections,
including the mention of particularly relevant values from
Tables 1--3.

\subsubsection{HD 37903} 
This star lies within the Orion molecular cloud and illuminates
the reflection nebula NGC 2023.  UV spectra indicate significant
quantities of vibrationally excited H$_2$ along the line of
sight (Meyer et al.\ 2001) and the {\it FUSE} spectrum confirms
this.  Meyer et al. (2001) concluded that the excitation was due
to UV fluorescence in a dense area of gas within 1 pc of the
star.  The line of sight $R_V$ is above average and may be even
larger within the dense material local to the star.  While the
UV extinction curve does not have a particularly unusual shape,
it does show smaller than normal extinction at all wavelengths
below 2500 \AA\ (Fitzpatrick \& Massa 1990).

\subsubsection{HD 38087}
This star illuminates the reflection nebula IC 435, with
estimates that about one-quarter or less of the line-of-sight
material is local to the star (Snow \& Witt 1989).  The line of
sight shows far less than normal far-UV extinction as well as a
significant shift of the 2175 \AA\ bump to longer wavelengths
(Fitzpatrick \& Massa 1990).  IR photometry indicates an
exceptionally large total-to-selective extinction ratio, $R_V$ =
5.57, consistent with the small far-UV extinction.  Also, Witt,
Bohlin, \& Stecher (1986) found evidence for scattering
associated with the 2175 \AA\ interstellar feature.  These data
strongly suggest that larger than normal dust grains have
developed in this presumably quiescent environment.  Enhanced
(but uncertain) depletions of manganese and zinc in the line of
sight also suggest grain mantle growth (Snow \& Witt 1989),
while the abundances of oxygen (Jensen, et al.\ 2005)
and nitrogen (Jensen, et al.\ 2007) are normal.

\subsubsection{HD 40893}
The extinction curve shows a relatively narrow and somewhat weak
2175 \AA\ bump and slightly enhanced far-UV extinction
(Jenniskens \& Greenberg 1993).  IR photometry also indicates an
abnormally small total-to-selective extinction ratio, $R_V$ =
2.46.
 
\subsubsection{HD 41117, HD 42087, HD 43384}
These stars all lie within the Gem OB1 association and sample a
roughly 2.5 $\times$ 4.5 degree field in the foreground of this
association.  Extinction curves are available for the first two
stars and are very similar and normal, and the other extinction
measures for all three stars are similar and near average.  HD
41117 and HD 42087 are Be stars and the uncertainties in the
photometric $R_V$ are large, but there is excellent agreement
with the other two techniques.

\subsubsection{HD 46056, HD 46202}
These stars lie within a quarter-degree of each other in open
cluster NGC 2244 within the Mon OB2 association and the lines of
sight show very similar extinction parameters.  In both cases,
$R_V$ is slightly smaller than average and the amount of far-UV
extinction is above average.  Also, the 2175 \AA\ bumps are
relatively weak and narrow.  As will be shown, the H$_2$
parameters are indeed nearly identical.

\subsubsection{HD 53367}
The line of sight to this star shows a relatively large amount
of reddening.  IR photometry suggests a smaller than normal
value of $R_V$, but this is a Be star and the value is very
uncertain.  There is no published extinction curve or
polarization data to verify the value of $R_V$.

\subsubsection{HD 147888}
This star is also known as $\rho$ Oph D and lies within the
$\rho$ Oph cloud complex, which is well known for having unusual
dust characteristics.  All three methods of assessing $R_V$
result in significantly above average values ($\sim$4--5) and
the extinction curve (Fitzpatrick \& Massa 1990) shows a
correspondingly small far-UV extinction.  This particular line
of sight has small reddening relative to other stars in this
complex, and when combined with the shallow far-UV extinction
curve, it was accessible to {\it FUSE} even though the star is
of relatively late spectral type and thus has a small UV flux.

\subsubsection{HD 149404}
This star is relatively bright optically and there have been
many studies of the material along the line of sight.  Despite
the significant number of optical and mm-wave observations, a UV
extinction curve has not been published for this star.  The
photometric value of $R_V$ is uncertain as HD 149404 is a Be
star.  However, polarization data also indicates a normal value
of $R_V$.

\subsubsection{HD 152236}
This star, also known as $\zeta^1$ Sco, is part of the Sco OB1
association.  The star is relatively bright given the amount of
extinction and thus provides one of several lines of sight with
$A_V$ $\approx$ 2 that were easy for FUSE to observe.  The UV
extinction curve is normal.  It is a Be star, but polarization
and extinction curve data confirm the apparently normal
photometric value of $R_V$.

\subsubsection{HD 164740}
Also known as Herschel 36, this star excites a compact H II
region within the Lagoon Nebula (M8, NGC 6523) known as the
Hourglass (Thackeray 1950, Wolff 1961).  The value of $R_V$ for
this line of sight, 5.36, is exceptionally large and
a correction for material in the foreground of M8 would give an
even larger $R_V$ (Hecht et al.\ 1982).  The unparametrized
far-UV extinction curve published by Hecht et al.\ (1982) is
also exceptionally shallow, and the new parametrization by
Fitzpatrick \& Massa (2007) gives a similar result.  That
characteristic allowed us to take the first UV spectrum of this
interesting line of sight at sufficient resolution and S/N to
investigate interstellar abundances despite the large total
extinction.  This spectrum shows a number of unusual and
interesting features which will be covered in a separate paper
(B.\ L.\ Rachford, in preparation, 2008), but we will give the
overall H$_2$ results here.

\subsubsection{HD 179406}
Also known as 20 Aql, this star shines through apparently
typical diffuse cloud material and lies toward the lower end of
the amounts of extinction covered in this study.  However, given
the small extinction, the abundances of carbonaceous molecules
are relatively large indicating that this line of sight may
sample a core of denser material (Hanson, et al.\ 1992).
Fitzpatrick \& Massa (2007) included this target in their
updated UV extinction curve parametrization work, which shows a
stronger than normal 2175 \AA\ bump, but is otherwise normal.

\subsubsection{HD 186994}
This star was primarily observed because it is bright and had
been previously observed with {\it Copernicus}.  Despite its
relatively large distance ($\sim$2500 pc) the quantity of
interstellar material along the line of sight is small.
Extinction information is limited for this star simply because
the small amount of reddening makes it difficult to properly
analyze the UV extinction curve or to determine $R_V$.  (In
fact, this star has been used as a lightly reddened comparison
star in extinction studies; e.g., Sasseen et al.\ 2002.)  The
H$_2$ column density observed with {\it Copernicus} was quite
small, log N(H$_2$) = 19.59 (Savage et al.\ 1977) and we confirm
this result.

\section{Observations and data analysis}
Table 4 gives information on our {\it FUSE} observations for the
new targets.  For all targets with multiple integrations, we
performed a shift-and-coadd procedure to combine the data for
each detector segment, but did not combine data from different
segments.  Note that in two cases (HD 179406 and HD 186994), we
obtained multiple observations of the target.  In both cases,
this was required due to the brightness of the targets as a
precaution against saturation of the detector.  A very short
preliminary observation was obtained from which the true flux
was determined before spending time on the full observation.
The preliminary observation represented a non-trivial fraction
of the total data, so we included both datasets in the final
co-added spectra.

We originally planned to use spectra that were uniformly
processed with version 2.4 (or later) of the CALFUSE pipeline,
including a revision of the column densities in Paper I that
were measured from earlier reductions.  However, in comparing
results from differing versions of CALFUSE, we have not seen
significant differences in the derived column densities,
presumably because the extremely broad profiles are not affected
by subtle changes in the algorithms.  In only a few cases did
the differences approach the value of the 1-$\sigma$
uncertainty.  Thus, we have not revised the older values, nor
used CALFUSE versions beyond 2.4 for the newer targets.

We used the same measurement techniques described in Paper I and
we will only give a very brief overview of those techniques.  We
fitted H$_2$ line profiles to the Lyman series (4,0), (2,0), and
(1,0) ro-vibrational bandheads, including the $J$ = 0, 1, and 2
lines.  Weaker lines from higher rotational states and other
interstellar species were fitted and removed from the broad
bandhead profiles.  Our profile fits included the effects of
overlapping wings from adjacent bandheads as the $J$ = 0 and 1
profiles are heavily damped and extremely broad.  The $J$ = 2
lines from these bandheads are often strong enough to show
damping wings as well, and these lines had to be included due to
blending with the broad $J$ = 0 and 1 profiles.

At most, we obtained 9 independent measurements of the $J$ =
0--2 column densities from the combinations of three
ro-vibrational bands and one to four detector segments covering
each band.  Poor data quality, severe stellar interference in a
particular H$_2$ band, and/or problems during the observation
itself occasionally led to fewer suitable combinations of bands
and detector segments.  The differences from one band/segment
combination to another were generally considerably larger than
the formal fit uncertainties, so we averaged each individual
measurement and used the sample standard deviation as our formal
uncertainty.

\section{Results and Discussion}

\subsection{Overall comments}
Table 5 gives our primary results for the lines of sight.  The
H$_2$ column densities were measured directly from the spectra
as already described.  We also include two derived quantities,
the hydrogen molecular fraction ($f_{\rm H2}$), and the
rotational temperature ($T_{01}$) for each line of sight.

The hydrogen molecular faction is defined in terms of the
molecular and atomic hydrogen column densities as

\begin{equation}
f_{\rm H2} = \frac{2N(H_2)}{2N(H_2) + N(H I)}.
\end{equation}

The rotational temperature (in Kelvin) is determined by applying the
Boltzmann equation to the ratio of the column densities in the
first two rotational states, yielding

\begin{equation}
\frac{N(1)}{N(0)} = 9e^{-171/T_{01}}.
\end{equation}

Solving for the temperature and expressing the column densities
as base-10 logarithms gives

\begin{equation}
T_{01} = \frac{74.0}{\log N(0) - \log N(1) + 0.954}.
\end{equation}

As in Paper I, we generally interpret the ratio between $N$(1)
(ortho-H$_2$) and $N$(0) (para-H$_2$) as the kinetic temperature
on the assumption that collisions with H$^{+}$ (and H3$^{+}$
when enough is present) dominate over other processes in
controlling this ratio.

There is evidence that under some circumstances ortho-H$_2$ can
be rapidly converted to para-H$_2$ on grains (Le Bourlot 2000).
In this case, the $N$(1)/$N$(0) ratio becomes lower, yielding a
lower rotational temperature.  This process appears to be more
likely at the low temperatures in the core of a relatively dense
and opaque cloud (Shaw et al.\ 2005).  The fact that we seem to
be seeing multiple diffuse clouds along the lines of sight
(Paper I) suggests that this process may not important for our
sample.

Shaw et al.\ (2005) have modeled the physical conditions in one
of our lines of sight from Paper I, HD 185418.  Their model
includes the Le Bourlot (2000) treatment of the ortho to para
conversion process.  Their derived kinetic temperature for the
gas was about 25\% lower than than our derived $T_{01}$ = 101
$\pm$ 14 K.  This line of sight has the largest $T_{01}$ of our
entire sample, and the gas density is relatively low ($n_H$ = 27
cm$^{-3}$, but Sonnentrucker et al.\ 2003 find an even lower
value of $n_H$ = 6.3 cm$^{-3}$).

Lacking a complete knowledge of whether or not our lines of
sight may have non-thermal $N$(1)/$N$(0) ratios, we will
generally assume that the measured $T_{01}$ values correlate
with kinetic temperature in some manner.  However, this
potential uncertainty should be kept in mind in the
interpretation of correlations of $T_{01}$ with other
parameters.

Overall, the second half of the sample shows column densities,
molecular fractions, and temperatures similar to those in Paper
I.  The primary difference is the presence of a sample of lines
of sight with large extinction, small N(H$_2$), and large $R_V$,
which we will discuss in detail in \S\ 4.3.

\subsection{General correlations with reddening}
We have updated several correlations with reddening from Paper I
using the new data.  In Figure 1 we present a plot of the H$_2$
column density versus color excess (unlike Figure 2 in Paper I,
we give $N(H_2)$ on a linear scale).  While this plot shows the
expected increase in H$_2$ as we look through more material, it
also reflects that our targets probe a wide variety of
environments in that we usually see a range of column densities
at a given color excess.

One particularly important relationship we can investigate with
the complete dataset is the gas-to-dust ratio.  Using {\it
Copernicus} data, Bohlin, et al.\ (1978) found
$N$(H$_{\rm tot}$)/$E(B-V)$ = 5.8 $\times$ 10$^{21}$ atoms
cm$^{-2}$ mag$^{-1}$.  In Figure 2, we present plots of
$N$(H$_{\rm tot}$) versus $E(B-V)$.  In the upper panel we
include the Bohlin et al.\ {\it Copernicus} data and our {\it
FUSE} data.  In the lower panel we only include our {\it FUSE}
data for lines of sight with direct determinations of $N$(H I)
and $N$(H$_2$), and also excluded Be stars as the color excesses
might be overstated due to the circumstellar emission.  This
should give the most homogeneous sub-sample possible that covers
a broad range in color excess.  When constrained to pass through
the origin, our error-weighted best-fit slope is (5.94 $\pm$
0.37) $\times$ 10$^{21}$ atoms cm$^{-2}$ mag$^{-1}$, essentially
identical to the {\it Copernicus} value for less reddened lines
of sight.  The solid line in both panels is the best fit, and it
is worth noting that many of the most discrepant {\it FUSE}
points in the upper panel are Be stars, which are removed in the
bottom panel.

For reference, we have also included an unconstrained fit in
these panels which furthermore does not include the
low-reddening point at $E(B-V)$ = 0.17 (HD 186994).  This
illustrates the significance of constraining the fit with points
at small reddening.  Clearly this line is not a good fit to the
low-extinction {\it Copernicus} points.  One possibility is that
the high-extinction sample has a different slope than the
low-extinction sample.  This would represent a difference in the
gas-to-dust ratio that might indicate a change in dust
properties in the clouds with higher extinction.  Such a
difference is not clearly seen in our data.  The slope of the
dashed line is (4.2 $\pm$ 1.7) $\times$ 10$^{21}$ atoms cm$^{-2}$
mag$^{-1}$, less than the value for the constrained fit, but
with large enough uncertainty to be consistent with the
constrained fit.

We performed the same analysis for the total visual extinction,
$A_V$, which is simply the product of $E(B-V)$ and $R_V$.
Figure 3 shows panels corresponding to the ones shown in Figure
2.  Again, we have fitted a line to our ``best'' sub-sample in
the lower panel, which yields $N$(H$_{\rm tot}$)/$A_V$ = (2.15
$\pm$ 0.14) $\times$ 10$^{21}$ atoms cm$^{-2}$ mag$^{-1}$.  This
value is nearly identical to a simple division of 5.94 $\times$
10$^{21}$ by the average $R_V$ = 2.93 for this sub-sample; the
latter value is slightly less than the galactic average of 3.1
(Draine 2003).  Again, excluding HD 186994 and not constraining
the fit to pass through the origin gives a slightly smaller
slope of (1.50 $\pm$ 0.43) $\times$ 10$^{21}$ atoms cm$^{-2}$
mag$^{-1}$.  Visually, there is a hint of a slope change in the
$A_V$ = 1.5--2.0 interval in the upper panel, but with so few
reliable points with $A_V$ $\geq$ 2, such a trend is not clear.

As Figure 4 shows (an update of Figure 5 in Paper I), there
appears to be a weak inverse correlation between rotational
temperature and reddening.  However, it is critical to note that
much of this trend disappears if we were to exclude the {\it
Copernicus} data points with $N$(H$_2$) $<$ 10$^{20}$ cm$^{-2}$.
In the region of overlap with {\it Copernicus} data, the {\it
FUSE} lines of sight show smaller average temperatures, while
the {\it FUSE} data for large reddening do not deviate from the
similar data for small reddening.  The mean $T_{01}$ for our
entire {\it FUSE} sample is 67 K with a standard deviation of 14
K.  This result is similar to other studies, including the
$<T_{01}>$ = 86 $\pm$ 20 K from the {\it FUSE} Galactic disk
survey (J.\ M.\ Shull et al., in preparation).  However, the
$T_{01}$ distribution of our sample is not normal, having a
general rise in frequency up to around 75--80 K, and just a few
stars with temperatures above that; the latter fact can easily
be seen in Figure 4.

In principle, the larger color excesses could be associated with
denser clouds, which in turn might be expected to show lower
temperatures.  In Paper I we noted a correlation between the
rotational temperature and the fractional abundance of the CN
radical, the latter of which is a good density indicator
(Federman, et al.\ 1984).  However, as we discussed in
Paper I, the lines of sight in the translucent sample seem to
mostly be made up of multiple diffuse clouds.  Thus, while the
slight trend is in the right direction to suggest that we
are probing somewhat denser clouds in the {\it FUSE} sample,
this conclusion is somewhat speculative.

Figure 5 (an update of Figure 7 in Paper I) shows a similar
pattern to Figure 1 in that once H$_2$ becomes self-shielded and
relatively abundant, the molecular fraction covers nearly the
entire possible range at any given reddening.  Again, part of
this may be a selection effect as the lines of sight were chosen
to sample a variety of environments.  However, it is notable
that even with the additional sample we have not found a line of
sight with $f_{\rm H2}$ $>$ 0.8.  In fact, as seen in Table 5,
the new lines of sight preferentially have relatively small
$f_{\rm H2}$ given the large extinctions.

\subsection{The high $R_V$ sample}
\subsubsection{Significance}
An important aspect of the completed translucent sample is that
we have determined the molecular hydrogen column densities for
several lines of sight with particularly large values of $R_V$
combined with larger $A_V$ and total hydrogen column densities
than previous samples.  In some cases, these lines of sight show
unusually small molecular fractions and are of particular
interest because of the relationship between dust parameters and
molecular hydrogen formation and destruction.

Snow (1983) first investigated the possibility that H$_2$
abundances may be inversely correlated with grain size.  Larger
grains, through the coagulation of small grains, provide less
surface area per unit dust mass.  With less surface area available,
the H$_2$ formation rate should be diminished.  Using the
compilation of {\it Copernicus} data of Bohlin et al.
(1978), Snow demonstrated that the mean molecular fraction in
the $\rho$ Oph cloud was a factor of 2.6 less than the
rest of the sample.

Cardelli (1988) explored the related issue of the relationship
between H$_2$ abundances and $R_V$.  The H$_2$ abundances again
came from Bohlin et al. (1978), while the values of
$R_V$ came from an analysis of IR photometry similar to that
which we have applied in the present work.

Cardelli's sample displayed an inverse relationship between the
H$_2$ to $A_V$ ratio and $R_V$, with a weaker, but similar
dependence of the hydrogen molecular fraction and $R_V$.  When
splitting the {\it Copernicus} lines of sight at $R_V$ = 3.5,
the high $R_V$ group had an average molecular fraction of
approximately a factor of 2.5 less than the low $R_V$ group.  As
$R_V$ is thought to be positively correlated with grain size (or
grain size distribution) this result provides further support
for the role of grain size in regulating the H$_2$ abundance.
Furthermore, the large values of $R_V$ are associated with
smaller than normal far-UV extinction, which allows more
photodissociating radiation to penetrate the interstellar clouds
and provides further reduction in the hydrogen molecular
fraction.  In fact, it appears that only the lines of sight with
the largest $R_V$ have extinction curves that deviate in a
consistent manner from average (Fitzpatrick \& Massa 2007).

In Paper I, we had minimal coverage of high $R_V$ lines of sight
in our {\it FUSE} sample, but the newly added targets improve
the situation.  In Figure 6 we show the relationship between
$f_{\rm H2}$ and $R_V$ for the {\it FUSE} and {\it Copernicus}
samples.  This provides a major update to Figure 8 in Paper I as
we have also included the {\it Copernicus} sample, using 2MASS
photometry to derive the values of $R_V$ as with the {\it FUSE}
sample.

Interestingly, using the Spearman rank correlation coefficient,
at best we only see an anti-correlation between molecular
fraction and $R_V$ at about the 1 $\sigma$ level whether or not
we restrict the sample to the best determinations of either
parameter.  A more significant trend (3 $\sigma$) appears when
considering the H$_2$ to $A_V$ ratio versus $R_V$ similar to
Cardelli's result, but as he discussed, this division works best
when one can make the assumptions of similar average densities
in the various clouds and that most of the extinction occurs in
the regions where H$_2$ is significantly present.  However, it
is not clear that this is an appropriate way to look at the {\it
FUSE} translucent sample.  In particular, much of the trend
disappears if we exclude lines of sight with $A_V$ $>$ 2.  As
discussed in Paper I, we believe that most of our lines of sight
are sampling multiple clouds that may be spread out in space,
and thus may have a variety of conditions.

We wish to consider in more detail the lines of sight with
$R_V$ $>$ 4, which are labeled in Figure 6, split equally
between the {\it FUSE} and {\it Copernicus} samples.  We will
ignore the {\it Copernicus} target HD 10516 ($\phi$ Per) due to
the very large uncertainty in $R_V$ and small extinction
($E(B-V)$ $\approx$ 0.2 and $A_V$ $\sim$ 1).  We will thus focus
our discussion on the five remaining lines of sight with high
$R_V$ and large extinction.

Detailed modeling of an individual line of sight is complex and
requires a large amount of data to constrain the model, although
this has been done for a few favorable lines of sight from Paper
I (Rachford et al.\ 2001; Sonnentrucker et al.\ 2003).  Barring
such an analysis, we can get a sense of the strength of the
local UV radiation field using the high $J$ lines of H$_2$, and
thus an estimate of the importance of photodissociation.  A
preliminary analysis of lines from $J$ = 2 up to the highest
observable levels in a line of sight yields column densities
that can be used to estimate the amount of radiative pumping to
these levels.  This has proven challenging in many of the
translucent lines of sight due primarily to data quality and the
resulting difficulty in measuring weak lines and properly
interpreting saturated lines.  This is particularly acute in
these heavily reddened lines of sight as in many cases the S/N
rapidly decreases with decreasing wavelength.  Also, the
strengths of other lines are typically larger than in less
reddened lines of sight which causes more interference with the
high-$J$ lines.  However, we have enough information for most of
these five lines of sight to use $N$(6) and $N$(7) as potential
indicators of the strength of the radiation field.  These lines
are typically weakly saturated and not as difficult to interpret
as the stronger lines from lower rotational states, when the
data quality permits their detection or the derivation of well
constrained upper limits.  For reference, the column densities
for HD 110432 (Rachford et al.\  2001) were log $N$(6) = 14.20
$\pm$ 0.20 and log $N$(7) = 13.25$^{+1.25}_{-1.00}$ for a
radiation field that was modeled as twice the strength of the
average curve of Draine (1978).  Particle density also plays a
significant role in controlling H$_2$ excitation, so we should
look at the high $J$ column densities as an indicator of the
strength of the radiation field, but not as a definitive
measurement.

\subsubsection{HD 38087}
HD 38087 has $f_{\rm H2}$ greater than half, the largest of the
group, but this value is uncertain because we do not have a
direct measurement of interstellar N(H I) due to the very late
spectral type.  Fitzpatrick \& Massa (1990) report log N(H I) =
21.48 from Lyman $\alpha$.  However, at spectral type B5 V, the
interstellar line is strongly contaminated or even dominated by
the stellar line (Shull \& van Steenberg 1985; Diplas \& Savage
1994).

Atomic hydrogen column density is highly correlated with the
strength of certain diffuse interstellar bands, including
$\lambda$5780 (S.\ D.\ Friedman et al.\ 2008, in preparation; see
Herbig 1993 or Welty et al.\ 2006 for similar correlations), and
applying this correlation to HD 38087 gives log N(H I) = 21.08,
similar to our quoted value and which would only reduce the
derived molecular fraction from 0.52 to 0.42.  Thus, it seems
likely that this line of sight is genuinely rich in molecules.

We have already noted in \S\ 2.3.2 that this line of sight
samples a quiescent environment where grain growth might be
expected to occur (Snow \& Witt 1989).  However, only about 25\%
of the line of sight material is thought to be in the reflection
nebula.  The extinction curve indicates that the interstellar
material should be relatively transparent to photodissociating
far-UV radiation -- if such radiation exists at the cloud
location.  HD 38087 itself is of rather late spectral type, so
it may not contribute significantly to the far-UV radiation
field at the location of the bulk of the line of sight material.
However, the {\it FUSE} spectrum clearly shows H$_2$ lines up to
the $J$ = 7 level and we derive a logarithmic column density of
about 15.0 in that level, a very large value that implies a
significant excitation mechanism.

It is important to note that even a ``large'' quantity of
excited H$_2$ corresponds to a very small fraction of the total
H$_2$ column density.  One simply needs a particularly high
excitation temperature to produce a relatively ``flat''
rotational distribution with a high percentage of the H$_2$ in
the excited states, while keeping the total column density of
this material several orders of magnitude below the 10$^{20-21}$
cm $^{-2}$ totals seen in the translucent lines of sight.  It
thus may be the case that the excited H$_2$ is produced in the
reflection nebula, while the bulk of the low-excitation H$_2$ is
found farther from the star where photodissociation is not as
important.

\subsubsection{HD 148184}
The other line of sight from the high $R_V$ sample with
significant molecular material is HD 148184 ($\chi$ Oph), with 
$f_{\rm H2}$ = 0.33 (Bohlin et al.\ 1978).
This star is about 5 degrees from the $\rho$ Oph grouping and
samples material from the general Sco-Oph cloud complex.  The
{\it Hipparcos} parallax is 6.21 $\pm$ 0.23 mas (van Leeuwen
2007), corresponding to a distance of 161 $\pm$ 6 pc, which
places it near the distant edge of the interstellar material.
The star itself has spectral type B2 IVpe and may contribute
significant UV radiation to the material in the line of sight,
nearly all of which is likely associated with the cloud complex.
However, in the overall sense this portion of the complex is not
as highly populated by B-type stars as in the immediate $\rho$
Oph area.

This star was observed with {\it Copernicus} at high resolution,
but with very limited wavelength coverage.  Frisch (1980)
reported upper limits for two $J$ = 6 lines, yielding an upper
limit of log $N$(6) $<$ 14.33.  Thus, there does not appear to
be an unusually large source of excitation. 

One final important consideration for this star is that it is
the only one of the five which is a Be star and thus the caveats
given in \S\ 2.2 apply.  As the large error bar in Figure 6
implies, we found the color excesses to be a poor fit to the
Martin \& Whittet (1990) relation.  There is no available
extinction curve for this star.  The wavelength of maximum
polarization is 0.55 $\mu$m (Coyne et al. 1974), corresponding
to $R_V$ = 3.08.  It is thus possible that this line of sight
should not be in the high $R_V$ group.

\subsubsection{The $\rho$ Oph cloud: $\rho$ Oph A \& D}
The $\rho$ Oph cloud has long been known as a location which
exhibits large $R_V$ and larger than normal dust grains
(Carrasco, et al.\ 1973; Whittet 1974).  The name $\rho$
Oph is applied to four stars within a few arcminutes of each
other labeled A through D.  The A component (HD 147933) is the
brightest star in both the visible and UV and was observed by
{\it Copernicus}.  The D component (HD 147888) is part of the
present {\it FUSE} sample.  Improved {\it Hipparcos} distances
to these stars (van Leeuwen 2007) have been used as part of a
study of the distribution of interstellar material in the cloud
by Snow et al.\ (2008).  These distances are 111$^{+12}_{-10}$ pc
for $\rho$ Oph A and 125$^{+14}_{-11}$ pc for $\rho$ Oph D.  The
overall analysis indicates that both stars are in front of the
denser material that is sampled by the more distant and more
heavily obscured line of sight to HD 147889 that we had hoped to
study with {\it FUSE} as noted in \S\ 2.1.

The uncertainties in the {\it Copernicus} measurements of
$N$(H$_2$) and $N$(H I) for $\rho$ Oph A are relatively large,
but still strongly indicate a small molecular fraction.
Unfortunately, the spectral type of $\rho$ Oph D (B5 V) is late
enough that the interstellar Ly$\alpha$ line will be severely
contaminated by the stellar line, as with HD 38087, thus $N$(H
I) is very uncertain.  Cartledge et al.\ (2004) indirectly
estimated a value log $N$(H$_{\rm tot}$) = 21.73 $\pm$ 0.09
based on measurements towards nearby $\rho$ Oph A.  For this
paper, we have used our preferred method for stars later than
spectral type B2, namely, applying the Bohlin et al. (1978)
$N$(H$_{\rm tot}$)/$E(B-V)$ = 5.8 $\times$ 10$^{21}$ atoms
cm$^{-2}$ mag$^{-1}$ value we discuss in \S\ 4.2.  This gives
log $N$(H$_{\rm tot}$) = 21.44, much smaller than the value for
$\rho$ Oph A, but in excellent agreement with the correlation
between $N$(H I) and the equivalent width of the $\lambda$5780
DIB discussed in \S\ 4.3.2.  It should be noted that $\rho$ Oph
A is well known as a line of sight with a larger than normal
gas-to-dust ratio (e.g.\ Bohlin et al.\ 1978).  For that reason,
our derived value of log $N$(H$_{\rm tot}$) for $\rho$ Oph D may
also be too low.  A larger value of $N$(H$_{\rm tot}$) would
produce an even smaller molecular fraction than the $f_{\rm H2}$
= 0.21 depicted in Figure 6.  Thus, it appears that the
molecular fractions toward both stars are relatively small given
the amount of reddening and extinction.

There are a number of B-type stars in the vicinity and the UV
radiation field is thought to be strong in this area despite the
lack of O-type stars (e.g., Kulesa et al. 2005).  Our
preliminary analysis of the high $J$ lines in the {\it FUSE}
spectrum of $\rho$ Oph D reveals no conclusive detections beyond
$J$ = 6, from which we derive an uncertain log $N$(6) = 14.3.  This
is comparable to that found toward HD 110432 (Rachford et al.
2001) which was modeled with a radiation field twice the
interstellar average curve of Draine (1978).

\subsubsection{Herschel 36}
As noted in \S\ 2.3.10, HD 164740, better known as Herschel 36,
lies within the Lagoon Nebula (M8) in a region with recent and
ongoing star formation.  The molecular fraction (0.03) is the
smallest known for a line of sight with $E(B-V)$ $>$ 0.3.  It is
believed that some of the intervening material lies close to the
star and is thus subject to the very strong far-UV radiation
field of the late O-type star.  In fact, our {\it FUSE} spectrum
shows extreme H$_2$ excitation, including numerous lines from
vibrationally excited states (B.\ L.\ Rachford, in preparation)
demonstrating the likelihood of significant H$_2$ lying close to
the star and the resultant exposure to a large far-UV flux.
Furthermore, the exceptionally small far-UV extinction will
allow the radiation from the O-type stars in M8 to influence a
greater distance along the line of sight.  Thus, while this line
of sight samples a complicated environment, it is plausible to
assume that the radiation field is contributing to the small
line of sight molecular fraction.

\subsubsection{Overall properties of this sample}
The 5 lines of sight were chosen for having evidence for larger
than normal grains, thus providing smaller than normal H$_2$
formation rates per unit dust mass.  Also, the far-UV extinction
is small which allows greater than normal penetration of the
photodissociating radiation.  However, there remains a large
range in molecular fractions that spans most of the total range
we see in the overall translucent sample.

In formation-destruction equilibrium, the hydrogen molecular
fraction will be proportional to the factors
controlling formation, which are density ($n_H$) and the
formation rate coefficient ($R$), and inversely proportional to
the far-UV radiation field which controls destruction.  For the
moment, we will focus on the last point.

There seems to be some evidence that the strength of the local
radiation field is responsible for this range, particularly when
considering the $\rho$ Oph cloud and Herschel 36.  However, the
line of sight toward HD 38087 is an exception.  As already
noted, the distribution of material along this line of sight
likely also plays a significant role.

Interestingly, these lines of sight have small H$_2$ rotational
temperatures.  In Figure 7 (an update of Figure 9 in Paper I) we
show the relationship between molecular fraction and rotational
temperature.  In Paper I, we highlighted the high $f_{\rm H2}$,
low $T_{01}$ lines of sight.  Much rarer are lines of sight with
low $f_{\rm H2}$ and low $T_{01}$.  In fact, $\rho$ Oph A and D
and Herschel 36 are the only lines of sight in the {\it
Copernicus}/{\it FUSE} sample with $f_{\rm H2}$ $<$ 0.2,
$T_{01}$ $\leq$ 60 K, and $E(B-V)$ $>$ 0.2 (or log N(H) $>$
21.1).

In contrast to the high $R_V$ lines of sight with small
molecular fraction and low temperature, there are numerous lines
of sight with normal or low values of $R_V$ that have small
molecular fractions and higher than normal temperatures,
particularly within the {\it Copernicus} sample.  Gas heating
due to grain electron photoemission (Draine 1978) may contribute
to this fact, although there is not a significant overall
relationship between $R_V$ and $T_{01}$.  The 13 {\it
Copernicus} lines of sight with $R_V$ $<$ 4 and $f_{\rm H2}$ $<$
0.2 all have $E(B-V)$ = 0.2--0.4, and all but one (HD 147165;
$T_{01}$ = 64 K) have rotational temperatures greater than the
average from our {\it FUSE} sample of 67 K.  Of the 4 {\it FUSE}
lines of sight with $R_V$ $<$ 4 and $f_{\rm H2}$ $<$ 0.2, only
one has $T_{01}$ $<$ 67 K (HD 152236; $T_{01}$ = 62 K).  Of
these 17 {\it Copernicus} and {\it FUSE} total lines of sight
none have $T_{01}$ as small as $\rho$ Oph A, $\rho$ Oph D, or
Herschel 36. 

It is clear from Figure 7 that these three lines of sight are
part of a small group that stand out from the rest of the
sample, going against the generally inverse relationship between
molecular fraction and temperature and lying in the bottom left
of the figure.  We would expect the cold lines of sight to
contain denser material and represent the expected environment
for the large grains.  But, for at least Herschel 36 and $\rho$
Oph D, the material is subject to considerable far-UV radiation,
contributing to the small molecular fractions.

In these cases, we may be seeing the effect of a broad
distribution of material.  Perhaps there is both ``cold''
diffuse material which contains most of the H$_2$ and is still
not dense enough to exhibit a level of self-shielding that would
allow a large molecular fraction, yet there is also a relatively
small amount of material closer to the hot stars that not only
has few molecules, but also considerable H$_2$ excitation.  In
particular, for Herschel 36 there seems to be a significant
velocity difference between the excited material and the
``cold'' material as indicated by the low-$J$ lines of H$_2$.
Thus, the cold material may be significantly in the foreground
of Herschel 36.

As previously noted, the radiation field is one of three factors
that influence the molecular fraction in the models of Browning
et al.\ (2003), along with formation rate coefficient and
density.  The models show that for the column densities
we are sampling in this paper, there can be degeneracy in the
molecular fractions that result from different combinations of
the three factors.  In particular, a small formation rate
coefficient can lower the molecular fraction in a manner similar
to an increase in the UV flux.

The total column densities in our sub-sample of 5 high $R_V$
lines of sight cover the range log $N$(H$_{\rm tot}$) $\approx$
21.2-22.0 ($N$ in cm$^{-2}$).  In this range, Browning et al.\
(2003) find that a UV radiation field 50 times the Galactic mean
reduces the molecular fraction by $\sim$5 orders of magnitude at
the low end, to factors of $\sim$2--3 at the high end.  A
reduction in the formation rate coefficient by a factor of 10
reduces the molecular fraction by $\sim$2 orders of magnitude at
the low end and by a minimal amount at the high end.  These two
factors in combination were required to match H$_2$ data for the
Small and Large Magellanic Clouds at levels of extinction
generally below those in our present sample.

The most likely variables that would change the formation rate
coefficient are grain size and temperature.  Since we have
limited this sub-sample to high $R_V$, differences in grain size
distribution have presumably been minimized.  If the
$N$(1)/$N$(0) rotational temperature is a meaningful indicator
of kinetic temperatures, our sub-sample is more or less
isothermal.

Density, the third factor considered in the Browning et al.\
(2003) models, is generally not as important within our range of
column densities for typical Galactic values of radiation field
and formation rate coefficient, particularly since the molecular
fractions are usually relatively large already.  When combined
with large radiation field or small formation rate coefficient,
density variations can cause a large spread in the resulting
small molecular fractions.

In conclusion, given the relatively large column densities in
the regime we are probing with our sample, it appears more
likely that large variations in radiation field are the dominant
cause for the variations in molecular fraction.  Formation rate
coefficients far below the Galactic average may contribute to
smaller molecular fractions in our regime, but it is more
certain that some of the clouds we are studying lie very close
to major UV sources that dramatically increase the local
radiation field.

\section{Summary}
We have completed the primary molecular hydrogen analysis for
the {\it FUSE} translucent lines of sight.  Total H$_2$ column
densities have been measured for a total of 38 lines of sight
with $A_V$ $\gtrsim$ 1 via profile fitting of transitions from
the lowest two vibrational levels which contain $\sim$99\% of
the material.  In addition, we have derived the H$_2$ molecular
fractions and rotational temperatures for the lines of sight and
considered various correlations between parameters.  In
particular, using these data we have found that the gas-to-dust
ratio ($N$(H$_{\rm tot}$)/$E(B-V)$) remains identical to that
found with {\it Copernicus} data out to $E(B-V)$ $\approx$ 1, a
factor of 2 farther than the previous determination.

These lines of sight were chosen to sample a wide variety of
environments, including those with unusual dust characteristics,
as dust grains are thought to provide the primary environment
for H$_2$ formation in these clouds.  An important consequence
of the updated sample is that we have much better coverage of
lines of sight with large total-to-selective extinction ratios
($R_V$) and smaller than normal far-UV extinction.  These
unusual characteristics are thought to indicate larger than
normal dust grains for which the grain area per unit mass will
be lower.  In the lines of sight with large grains, we still see
a large range in molecular fraction and can mostly attribute
this to a range in the strength of the local interstellar far-UV
radiation field, perhaps enhanced by material being widely
distributed along the line of sight and/or by variations in the
H$_2$ formation rate coefficient.  Overall, we do not see a
statistically significant trend of decreasing molecular fraction
with increasing $R_V$.

As our new lines of sight all have molecular fractions $f_{\rm
H2}$ $\lesssim$ 0.5, our conclusions regarding the presence of
truly ``translucent clouds'' with $f_{\rm H2}$ near unity are
unchanged.  Namely, the lines of sight in our survey are
primarily sampling multiple clouds without a high-extinction core
that is dominated by molecules.

Work is still ongoing to fully utilize the {\it FUSE} data for
the translucent sample, including a survey of the HD molecule
(Snow et al.\ 2008), and a detailed analysis
of the line of sight to Herschel 36 (B.\ L.\ Rachford 2008, in
preparation).

With the official end of the {\it FUSE} mission in 2007, direct
far-UV measurements of another significant sample of reddened
lines of sight will have to wait until another mission is
launched.  One of the goals of the planned Hubble
Service Mission 4 in late 2008 will be to install the Cosmic
Origins Spectrograph (COS) in the Hubble Space Telescope.  This
instrument is primarily designed to observe UV wavelengths
longward of 1150 \AA\ at high sensitivity and similar resolution
to that of {\it FUSE}.  This will provide access to lines from
numerous atomic and molecular species in more heavily reddened
lines of sight than any previous mission.  However, the
combination of the {\it HST} optics and {\it COS} may have
enough residual sensitivity at 1108 \AA\ to sample the longest
wavelength (0,0) ro-vibrational bandhead of H$_2$.  These
potential observations of H$_2$ would occur at much lower
resolution than {\it FUSE} ($R$ $\sim$ 3000) and will thus be
more difficult to analyze.  However, this might provide a
significant constraint on molecular fractions for very heavily
reddened lines of sight that were not accessible with {\it
FUSE}, possibly revealing true ``translucent clouds'' with
$f_{\rm H2}$ $\approx$ 1.

\acknowledgments
We thank the referee for useful comments.  This work is based on
data obtained for the Guaranteed Time Team by the NASA-CNES-CSA
{\it FUSE} mission operated by the Johns Hopkins University.
Financial support to U.S. participants has been provided by NASA
contract NAS5-32985 and NASA grant NNX08AC14G.  This research
has made use of the SIMBAD database, operated at CDS,
Strasbourg, France.

{\it Facilities:} \facility{FUSE (LWRS,MDRS)}

\clearpage

\begin{deluxetable}{ccccclc}
\tablecaption{Target list}
\tablewidth{0pt}
\tablehead{
  \colhead{Star} & \colhead{$\ell$} & \colhead{$b$} & \colhead{Assoc.}
& \colhead{$V$} & \colhead{MK type}
}
\startdata
HD \phn37903           &206.85 &   $-$16.54 &Orion   &\phn7.83 &B1.5 V   \\
HD \phn38087           &207.07 &   $-$16.26 &Orion?  &\phn8.30 &B5 V     \\
HD \phn40893           &180.09 &\phn$+$4.34 &        &\phn8.90 &B0 IV    \\
HD \phn41117           &189.65 &\phn$-$0.86 &Gem OB1 &\phn4.63 &B2 Iae   \\
HD \phn42087           &187.79 &\phn$+$1.77 &Gem OB1 &\phn5.75 &B2.5Ibe  \\
HD \phn43384           &187.99 &\phn$+$3.53 &Gem OB1 &\phn6.25 &B3 Ib    \\
HD \phn46056           &206.34 &\phn$-$2.25 &Mon OB2 &\phn8.16 &O8 V     \\
HD \phn46202           &206.31 &\phn$-$2.00 &Mon OB2 &\phn8.19 &O9 V     \\
HD \phn53367           &223.71 &\phn$-$1.90 &        &\phn6.96 &B0 IVe   \\
HD 147888              &353.65 &   $+$17.71 &Sco-Oph &\phn6.74 &B5 V     \\
HD 149404              &340.54 &\phn$+$3.01 &        &\phn5.47 &O9 Iae   \\
HD 152236              &343.03 &\phn$+$0.87 &Sco OB1 &\phn4.73 &B1 Ia+pe \\
HD 164740        &\phn\phn5.97 &\phn$-$1.17 &M8         &10.30 &O7.5 V   \\
HD 179406           &\phn28.23 &\phn$-$8.31 &        &\phn5.34 &B3 V     \\
HD 186994           &\phn78.62 &   $+$10.86 &        &\phn7.50 &B0 III   \\
\enddata
\end{deluxetable}

\begin{deluxetable}{cccccccccccc}
\tablecaption{Extinction parameters for Paper I sample and
present sample\tablenotemark{a}}
\tabletypesize{\footnotesize}
\tablewidth{0pt}
\tablehead{
& & \multicolumn{4}{c}{$R_V$} \\
\cline{3-7} \\
  \colhead{Star} & \colhead{$E(B-V)$}
& \colhead{Phot.\tablenotemark{b}}
& \colhead{Polar.\tablenotemark{c}} & \colhead{Ref}
& \colhead{E.C.\tablenotemark{d}}
& \colhead{Adopted} & \colhead{$A_V$}
}
\startdata
BD $+$31$^{\circ}$ 643 &0.85 &3.13$\pm$0.30 &3.75 &1 &3.54  &3.13$\pm$0.30 &2.66$\pm$0.26 \\
HD \phn24534        &0.59 &1.70$\pm$1.00  &3.47 &2   &      &3.47$\pm$0.30 &2.05$\pm$0.18\\
HD \phn27778        &0.37 &2.72$\pm$0.30  &     &    &      &2.72$\pm$0.30 &1.01$\pm$0.11\\
HD \phn62542        &0.35 &2.83$\pm$0.30  &3.27 &3   &2.14  &2.83$\pm$0.30 &0.99$\pm$0.14\\
HD \phn73882        &0.70 &3.37$\pm$0.30  &3.51 &3   &2.93  &3.37$\pm$0.30 &2.36$\pm$0.23\\
HD \phn96675        &0.30 &3.85$\pm$0.30  &2.80 &4   &4.02  &3.85$\pm$0.30 &1.16$\pm$0.15\\
HD 102065           &0.17 &               &     &    &2.89  &2.89$\pm$0.30 &0.49$\pm$0.10\\
HD 108927           &0.22 &3.14$\pm$0.30  &     &    &3.73  &3.14$\pm$0.30 &0.69$\pm$0.11\\
HD 110432           &0.51 &3.95$\pm$0.60  &3.30 &5   &      &3.95$\pm$0.60 &2.02$\pm$0.33\\
HD 154368           &0.78 &3.00$\pm$0.30  &     &    &3.14  &3.00$\pm$0.30 &2.34$\pm$0.25\\
HD 167971           &1.08 &3.17$\pm$1.00  &     &    &      &3.17$\pm$1.00 &3.42$\pm$1.08\\
HD 168076           &0.78 &3.55$\pm$0.30  &3.19 &6   &3.62  &3.55$\pm$0.30 &2.77$\pm$0.26\\
HD 170740           &0.48 &2.71$\pm$0.30  &3.08 &5   &      &2.71$\pm$0.30 &1.30$\pm$0.17\\
HD 185418           &0.50 &2.32$\pm$0.30  &     &    &3.98  &2.32$\pm$0.30 &1.16$\pm$0.17\\
HD 192639           &0.66 &2.84$\pm$1.00  &     &    &      &2.84$\pm$1.00 &1.87$\pm$0.67\\
HD 197512           &0.32 &2.35$\pm$0.30  &     &    &2.56  &2.35$\pm$0.30 &0.75$\pm$0.12\\
HD 199579           &0.37 &2.95$\pm$1.00  &     &    &2.74  &2.95$\pm$0.30 &1.09$\pm$0.14\\
HD 203938           &0.74 &2.91$\pm$0.30  &     &    &3.00  &2.91$\pm$0.30 &2.15$\pm$0.24\\
HD 206267           &0.53 &2.67$\pm$0.30  &     &    &      &2.67$\pm$0.30 &1.41$\pm$0.18\\
HD 207198           &0.62 &2.42$\pm$1.00  &2.30 &7   &2.66  &2.42$\pm$0.30 &1.50$\pm$0.20\\
HD 207538           &0.64 &2.25$\pm$0.30  &2.23 &3   &      &2.25$\pm$0.30 &1.44$\pm$0.20\\
HD 210121           &0.40 &2.08$\pm$0.30  &2.13 &8   &2.01  &2.08$\pm$0.30 &0.83$\pm$0.14\\
HD 210839           &0.57 &2.78$\pm$0.30  &2.86 &9   &      &2.78$\pm$0.30 &1.58$\pm$0.19\\
\\ 
HD \phn37903        &0.35 &3.67$\pm$0.30  &3.89 &3   &3.90  &3.67$\pm$0.30 &1.28$\pm$0.15\\
HD \phn38087        &0.29 &5.57$\pm$0.30  &3.06 &3   &4.48  &5.57$\pm$0.30 &1.61$\pm$0.19\\
HD \phn40893        &0.46 &2.46$\pm$0.30  &     &    &3.18  &2.46$\pm$0.30 &1.13$\pm$0.16\\
HD \phn41117        &0.45 &2.74$\pm$1.00  &3.02 &5   &2.89  &2.74$\pm$0.30 &1.23$\pm$0.16\\
HD \phn42087        &0.36 &3.06$\pm$1.00  &3.08 &5   &2.78  &3.08$\pm$0.30 &1.10$\pm$0.14\\
HD \phn43384        &0.58 &3.06$\pm$0.30  &2.97 &5   &      &3.06$\pm$0.30 &1.78$\pm$0.20\\
HD \phn46056        &0.50 &2.60$\pm$0.30  &     &    &2.81  &2.60$\pm$0.30 &1.30$\pm$0.17\\
HD \phn46202        &0.49 &2.83$\pm$0.30  &     &    &2.79  &2.83$\pm$0.30 &1.39$\pm$0.17\\
HD \phn53367        &0.74 &2.38$\pm$1.00  &     &    &      &2.38$\pm$1.00 &1.76$\pm$0.74\\
HD 147888           &0.47 &4.06$\pm$0.30  &3.82 &3   &4.93  &4.06$\pm$0.30 &1.91$\pm$0.19\\
HD 149404           &0.68 &3.28$\pm$0.60  &3.08 &5   &      &3.28$\pm$0.60 &2.23$\pm$0.42\\
HD 152236           &0.68 &3.29$\pm$1.00  &3.25 &5   &2.82  &3.29$\pm$0.30 &2.24$\pm$0.23\\
HD 164740           &0.87 &5.36$\pm$0.30  &3.75 &5   &      &5.36$\pm$0.30 &4.66$\pm$0.31\\
HD 179406           &0.33 &2.86$\pm$0.30  &2.86 &5   &      &2.86$\pm$0.30 &0.94$\pm$0.13\\
HD 186994           &0.17 &               &     &    &      &3.10$\pm$0.30 &0.53$\pm$0.13\\
\enddata
\tablenotetext{a}{Revised values for Paper I targets are given
first, followed by the values for the new targets}
\tablenotetext{b}{Derived using 2MASS photometry}
\tablenotetext{c}{Derived from the wavelength of maximum polarization}
\tablenotetext{d}{Derived from the linear far-UV rise in the
extinction curve based on the $c_2$ parameters given in Table 3.}
\tablerefs{(1) Andersson \& Wannier 2000;
  (2) Roche et al. 1997; (3) Martin, Clayton, \& Wolff 1999;
  (4) Whittet et al. 1994; (5) Serkowski, Mathewson, Ford 1975;
  (6) Orsatti, Vega, \& Marraco 2000; (7) Anderson et al. 1996;
  (8) Larson, Whittet, \& Hough 1996; (9) McDavid 2000
}
\end{deluxetable}

\begin{deluxetable}{cccccccc}
\tablecaption{Extinction curve parameters\tablenotemark{a}}
\tablewidth{0pt}
\tablehead{
  \colhead{Target} & \colhead{$\lambda_0^{-1}$} & \colhead{$\gamma$}
& \colhead{c$_1$} & \colhead{c$_2$} & \colhead{c$_3$} & \colhead{c$_4$}
& \colhead{Ref.} \\
& \colhead{($\mu$m$^{-1}$)} & \colhead{($\mu$m$^{-1}$)}
}
\startdata
HD \phn37903 &4.615 &1.045    &\phm{$-$}0.965    &0.384    &3.300    &0.440    &1 \\
HD \phn38087 &4.563 &1.026    &\phm{$-$}1.137    &0.230    &4.508    &0.311    &1 \\
HD \phn40893 &4.591 &0.83\phn &\phm{$-$}0.26\phn &0.66\phn &3.13\phn &0.55\phn &2 \\
HD \phn41117 &4.621 &0.97\phn &$-$0.38\phn       &0.81\phn &3.64\phn &0.56\phn &2 \\
HD \phn42087 &4.636 &1.05\phn &$-$1.28\phn       &0.87\phn &4.36\phn &0.53\phn &2 \\
HD \phn43384 &                                                                    \\
HD \phn46056 &4.611 &0.932    &$-$0.527          &0.857    &3.032    &0.541    &1 \\
HD \phn46202 &4.599 &0.842    &$-$0.348          &0.864    &2.542    &0.515    &1 \\
HD \phn53367 &                                                                    \\
HD 147888    &4.587 &1.022    &\phm{$-$}1.611    &0.133    &3.823    &0.339    &1 \\
HD 149404    &                                                                    \\
HD 152236    &4.622 &1.06\phn &$-$0.51\phn       &0.85\phn &3.71\phn &0.38\phn &2 \\
HD 164740    &                                                                    \\
HD 179406    &                                                                    \\
HD 186994    &                                                                    \\

\enddata
\tablenotetext{a}{In the parameterization scheme of Fitzpatrick \& Massa 1990}
\tablerefs{(1) Fitzpatrick \& Massa 1990; (2) Jenniskens \& Greenberg 1993}
\end{deluxetable}

\begin{deluxetable}{cclccc}
\tablecaption{{\it FUSE} observations}
\tablewidth{0pt}
\tablehead{
  \colhead{Target} & \colhead{{\it FUSE} data ID} & \colhead{Date} &
  \colhead{N$_{\rm int}$\tablenotemark{a}} &
  \colhead{t$_{\rm int}$\tablenotemark{b}} &
  \colhead{S/N\tablenotemark{c}} \\
& & & \colhead{(ksec)} & 
}
\startdata
HD \phn37903 &P1160601 &2001 Oct 18 &\phn4 &\phn4.0 &    20.4 \\
HD \phn38087 &P1160701 &2001 Oct 18 &\phn4 &\phn4.0 &    16.2 \\
HD \phn40893 &P2160101 &2001 Oct 14 &\phn3 &\phn7.1 &    12.7 \\
HD \phn41117 &P2160201 &2004 Mar 03 &\phn1 &\phn0.1 & \phn2.0 \\
HD \phn42087 &P2160301 &2001 Oct 15 &\phn6 &\phn2.9 &    12.1 \\
HD \phn43384 &P2160401 &2001 Oct 15 &\phn6 &\phn8.1 & \phn1.8 \\
HD \phn46056 &P2160901 &2003 Jan 25 &\phn3 &\phn7.1 & \phn8.1 \\
HD \phn46202 &P2161001 &2001 Oct 16 &\phn2 &\phn5.0 &    10.0 \\
HD \phn53367 &P1161101 &2001 Oct 26 &\phn7 &11.4    & \phn9.9 \\
HD 147888    &P1161501 &2003 Aug 21 &23    &12.2    &    12.8 \\
HD 149404    &P1161701 &2001 Aug 07 &38    &17.9    &    28.9 \\
HD 152236    &P1161801 &2001 Aug 08 &\phn4 &\phn4.6 & \phn6.1 \\
HD 164740    &P1162001 &2000 Aug 30 &\phn3 &\phn5.9 & \phn9.0 \\
HD 179406    &P2160701 &2001 Apr 27 &\phn1 &\phn0.1 & \phn3.7 \\
HD 179406    &P2160702 &2002 Jun 11 &\phn3 &\phn1.0 &    14.4 \\
HD 186994    &P2160801 &2001 Jul 02 &\phn1 &\phn0.1 & \phn6.3 \\
HD 186994    &P2160802 &2001 Sep 07 &\phn1 &\phn0.4 &    17.2 \\
\enddata
\tablenotetext{a}{Number of integrations}
\tablenotetext{b}{Total integration time}
\tablenotetext{c}{Average per-pixel S/N for a 1 \AA\ region of the LiF 1A
spectrum near 1070 \AA , between the Lyman (3,0) and (2,0) bandheads of H$_2$.
One resolution element corresponds to about 9 pixels.}
\end{deluxetable}

\begin{deluxetable}{cccccccccc}
\tablecaption{Molecular and atomic hydrogen parameters}
\tabletypesize{\footnotesize}
\tablewidth{0pt}
\tablehead{
  \colhead{Target} & \colhead{Bands} & \colhead{log $N$(H$_2$)}
& \colhead{log $N$(0)}
& \colhead{log $N$(1)} & \colhead{T$_{01}$} & \colhead{log $N$(H I)}
& \colhead{Ref} & \colhead{$f_{\rm H2}$} \\
& & \colhead{($N$ in cm$^{-2}$)} & \colhead{($N$ in cm$^{-2}$)} & \colhead{($N$ in cm$^{-2}$)}
& \colhead{(K)} &  \colhead{($N$ in cm$^{-2}$)} & &
}
\startdata
HD \phn37903  &\phn6 &20.92$\pm$0.06 &20.68$\pm$0.07 &20.54$\pm$0.05 &\phn68$\pm$\phn7 &21.17$\pm$0.10 &1 &0.53$\pm$0.09 \\
HD \phn38087  &\phn7 &20.64$\pm$0.07 &20.39$\pm$0.08 &20.29$\pm$0.05 &\phn70$\pm$\phn8 &20.91$\pm$0.30 &2 &0.52$\pm$0.20 \\
HD \phn40893  &\phn9 &20.58$\pm$0.05 &20.27$\pm$0.05 &20.28$\pm$0.05 &\phn78$\pm$\phn8 &21.50$\pm$0.10 &3 &0.19$\pm$0.06 \\
HD \phn41117  &\phn2 &20.69$\pm$0.10 &20.51$\pm$0.10 &20.22$\pm$0.10 &\phn59$\pm$\phn8 &21.40$\pm$0.15 &4 &0.28$\pm$0.13 \\
HD \phn42087  &\phn7 &20.52$\pm$0.12 &20.31$\pm$0.12 &20.11$\pm$0.12 &\phn64$\pm$11    &21.39$\pm$0.11 &1 &0.21$\pm$0.10 \\
HD \phn43384  &\phn2 &20.87$\pm$0.14 &20.59$\pm$0.10 &20.54$\pm$0.18 &\phn74$\pm$16    &21.27$\pm$0.30 &5 &0.44$\pm$0.24 \\
HD \phn46056  &\phn9 &20.68$\pm$0.06 &20.40$\pm$0.06 &20.35$\pm$0.06 &\phn74$\pm$\phn8 &21.38$\pm$0.14 &1 &0.29$\pm$0.09 \\
HD \phn46202  &\phn9 &20.68$\pm$0.06 &20.38$\pm$0.07 &20.38$\pm$0.07 &\phn78$\pm$\phn9 &21.58$\pm$0.15 &1 &0.20$\pm$0.09 \\
HD \phn53367  &\phn9 &21.04$\pm$0.05 &20.89$\pm$0.04 &20.52$\pm$0.07 &\phn56$\pm$\phn4 &21.32$\pm$0.30 &2 &0.51$\pm$0.19 \\
HD 147888     &\phn7 &20.47$\pm$0.05 &20.39$\pm$0.04 &19.71$\pm$0.10 &\phn45$\pm$\phn4 &21.44$\pm$0.30 &5 &0.18$\pm$0.14 \\
HD 149404     &\phn9 &20.79$\pm$0.04 &20.60$\pm$0.03 &20.34$\pm$0.05 &\phn61$\pm$\phn4 &21.40$\pm$0.15 &1 &0.33$\pm$0.09 \\
HD 152236     &\phn1 &20.73$\pm$0.12 &20.53$\pm$0.12 &20.29$\pm$0.12 &\phn62$\pm$10    &21.77$\pm$0.15 &1 &0.15$\pm$0.08 \\
HD 164740     &\phn1 &20.19$\pm$0.12 &19.92$\pm$0.12 &19.86$\pm$0.12 &\phn60$\pm$10    &21.95$\pm$0.15 &6 &0.03$\pm$0.02 \\
HD 179406     &\phn4 &20.73$\pm$0.07 &20.55$\pm$0.07 &20.26$\pm$0.08 &\phn59$\pm$\phn6 &21.23$\pm$0.15 &7 &0.39$\pm$0.12 \\
HD 186994     &\phn9 &19.59$\pm$0.04 &19.18$\pm$0.06 &19.37$\pm$0.03 &\phn97$\pm$10    &20.90$\pm$0.15 &8 &0.09$\pm$0.04 \\

\enddata
\tablerefs{(1) Diplas \& Savage 1994, Ly$\alpha$;
(2) Jensen et al. 2007, $N$(H I) = 5.8$\times$10$^{21}$$E(B-V)$ -- 2$N$(H$_2)$;
(3) Jensen et al. 2007, Ly$\alpha$;
(4) Present work, Ly$\alpha$;
(5) Present work, $N$(H I) = 5.8$\times$10$^{21}$$E(B-V)$ -- 2$N$(H$_2)$;
(6) Fitzpatrick \& Massa 1990, Ly$\alpha$;
(7) Hansen et al. 1992, Ly$\alpha$;
(8) Bohlin et al. 1978, Ly$\alpha$}
\end{deluxetable}

\clearpage
\begin{figure}
\plotone{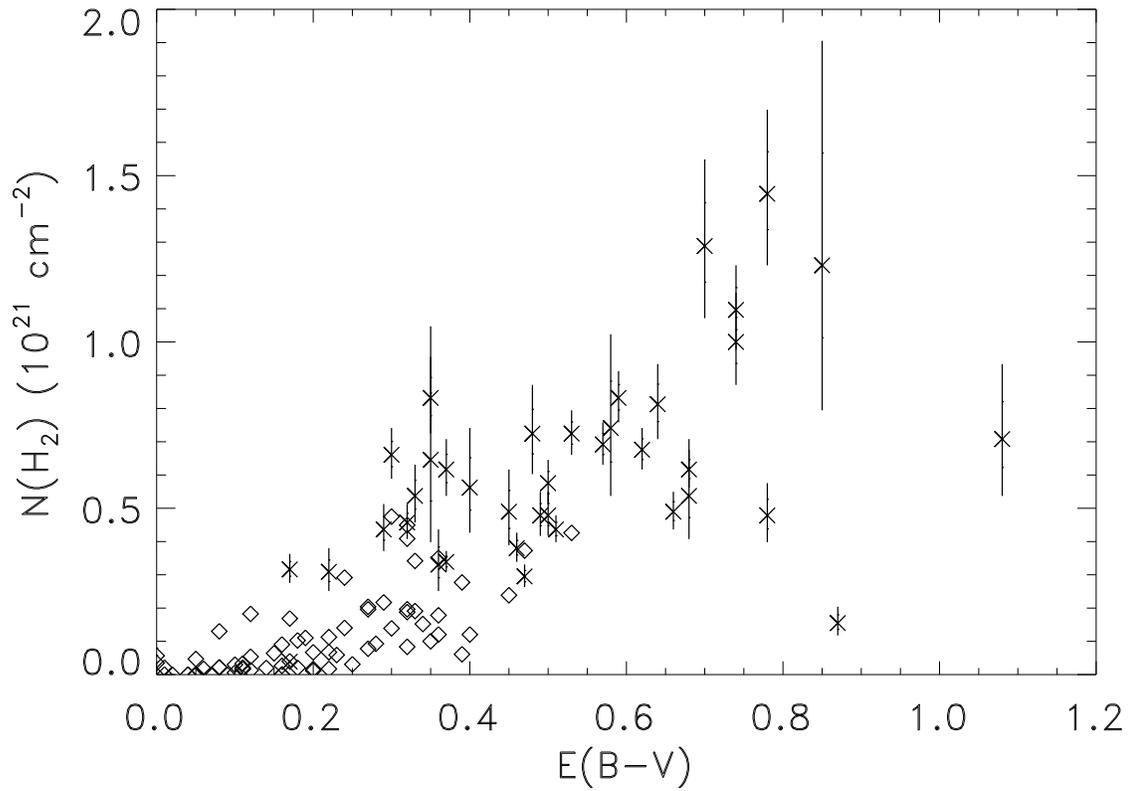}
\caption{Molecular hydrogen column density versus color excess.
Crosses: {\it FUSE}; diamonds: {\it Copernicus}.  Typical H$_2$
uncertainties for the {\it Copernicus} data points ($\sim$25\%)
are slightly larger than for the {\it FUSE} data points.}
\end{figure}

\clearpage
\begin{figure}
\epsscale{.95}
\plotone{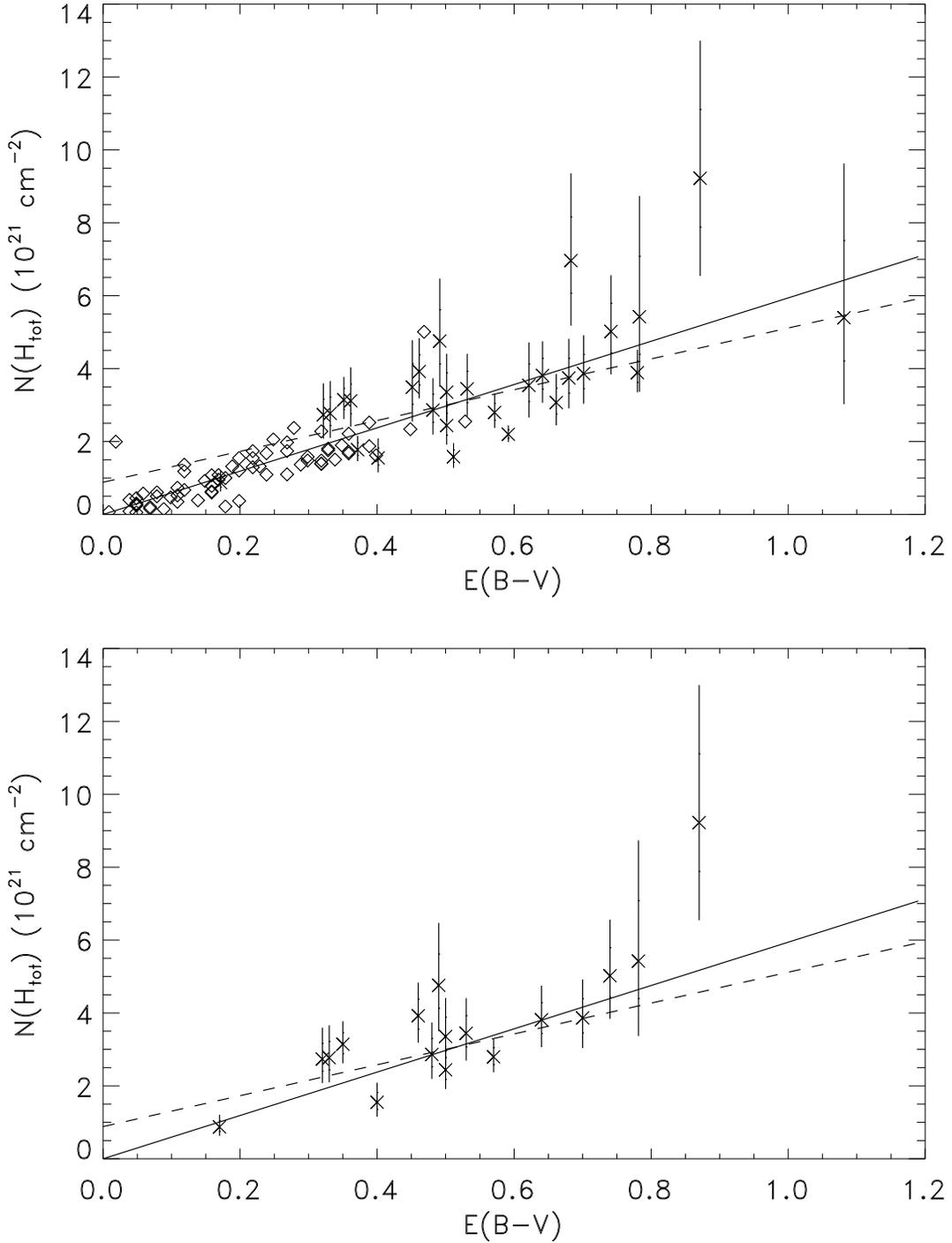}
\caption{Total hydrogen column density versus color excess.
Symbols as in Figure 1.  Top panel: All data points.  Bottom
panel: Only {\it FUSE} points with reliable H I and H$_2$
measurements toward non-Be stars.  Solid line in both panels is
the best fit to the data in the lower panel, constrained to pass
through the origin.  The dashed line is an unconstrained fit
that does not include the point at $E(B-V)$ = 0.17 (HD 186994).
Typical $N(H_{\rm tot})$ errors for the {\it Copernicus}
data points ($\sim$30\%) are slightly larger than for the {\it
FUSE} data points.}
\end{figure}

\clearpage
\begin{figure}
\plotone{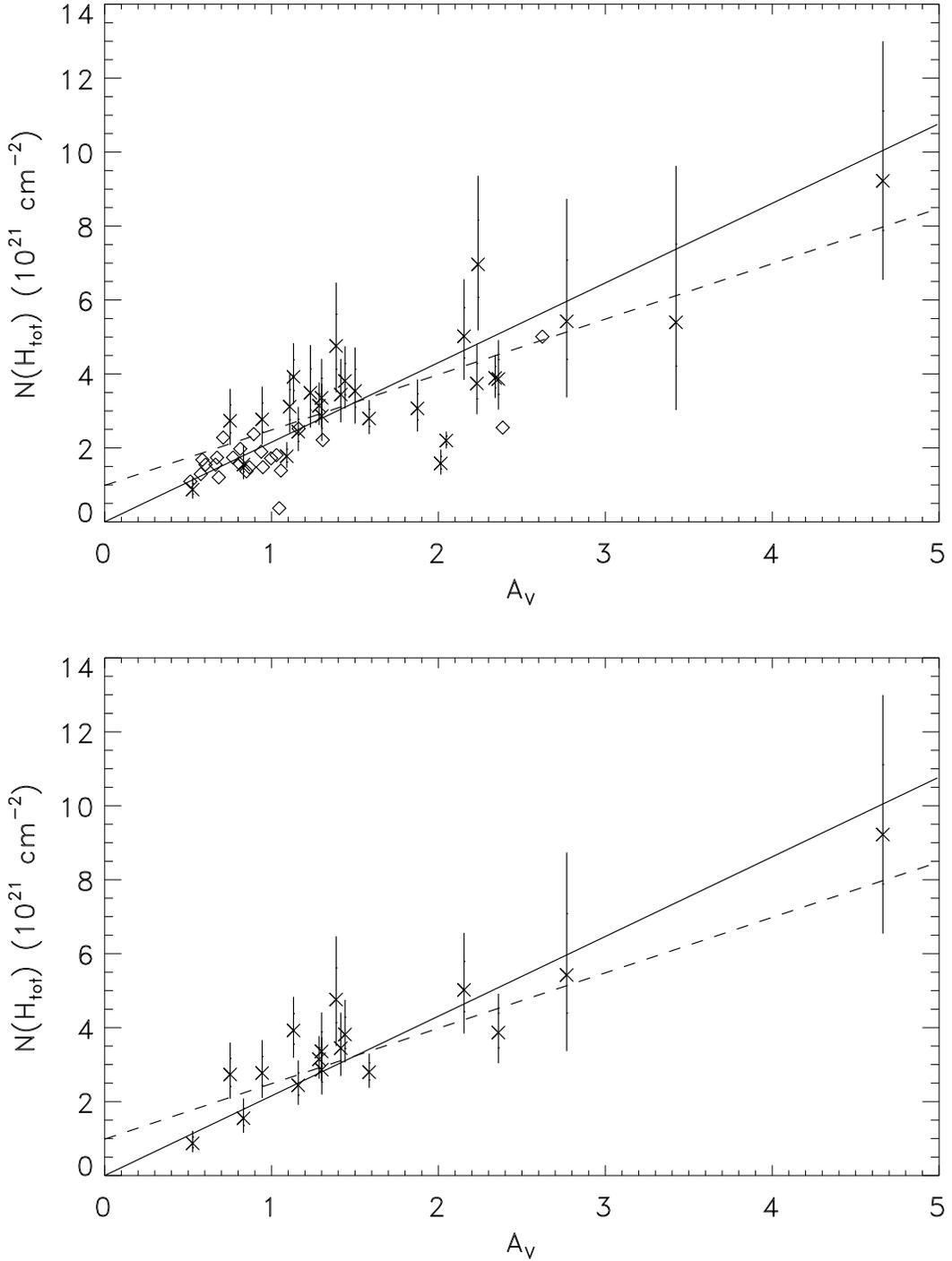}
\caption{Total hydrogen column density versus total visual
extinction.  Symbols as in Figure 1.  Top panel: All data
points.  Bottom panel: Only {\it FUSE} points with reliable H I
and H$_2$ measurements toward non-Be stars.  The solid line in
both panels is the best fit to the data in the lower panel,
constrained to pass through the origin.  The dashed line is an
unconstrained fit that does not include the point at $A_V$ =
0.53 (HD 186994).  Typical $N(H_{\rm tot})$ errors for the {\it
Copernicus} data points ($\sim$30\%) are slightly larger than
for the {\it FUSE} data points.}
\end{figure}

\clearpage
\begin{figure}
\epsscale{1}
\plotone{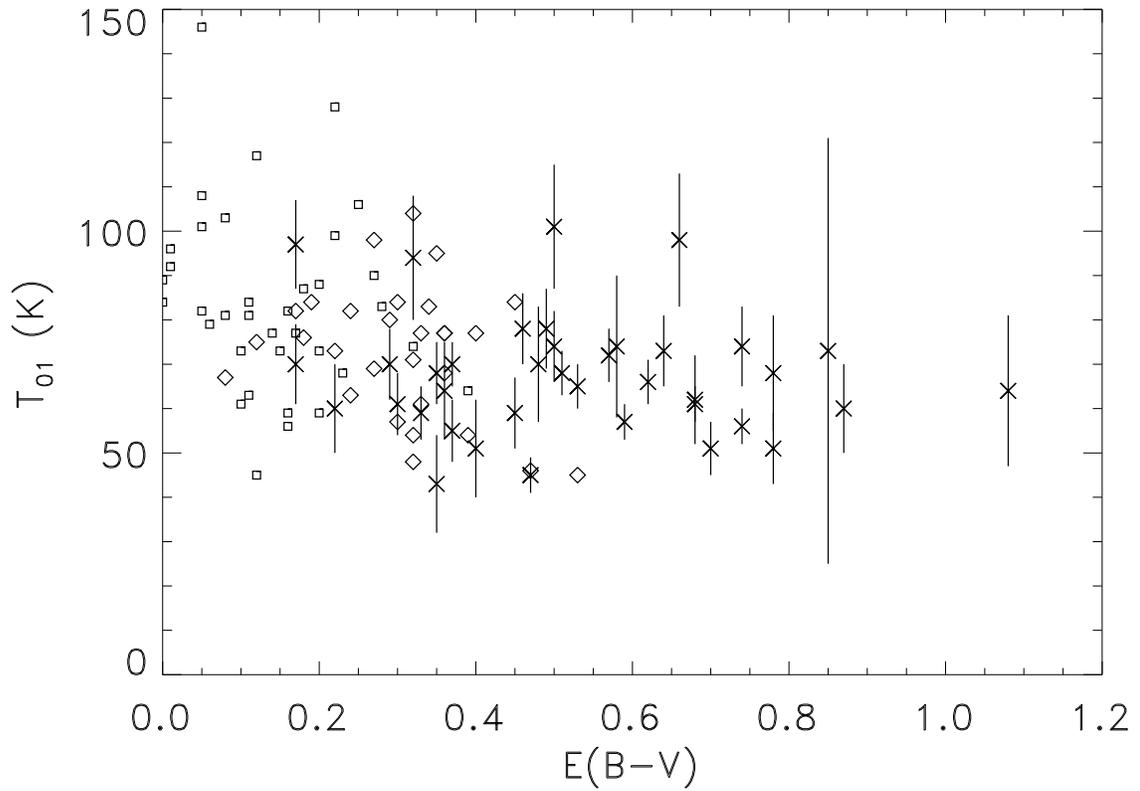}
\caption{Rotational temperature versus color excess.  Crosses:
{\it FUSE}; diamonds: {\it Copernicus} points with $N$(H$_2$) $>$
10$^{20}$ cm$^{-2}$; squares: {\it Copernicus} points with
$N$(H$_2$) $<$ 10$^{20}$ cm$^{-2}$.}
\end{figure}

\clearpage
\begin{figure}
\plotone{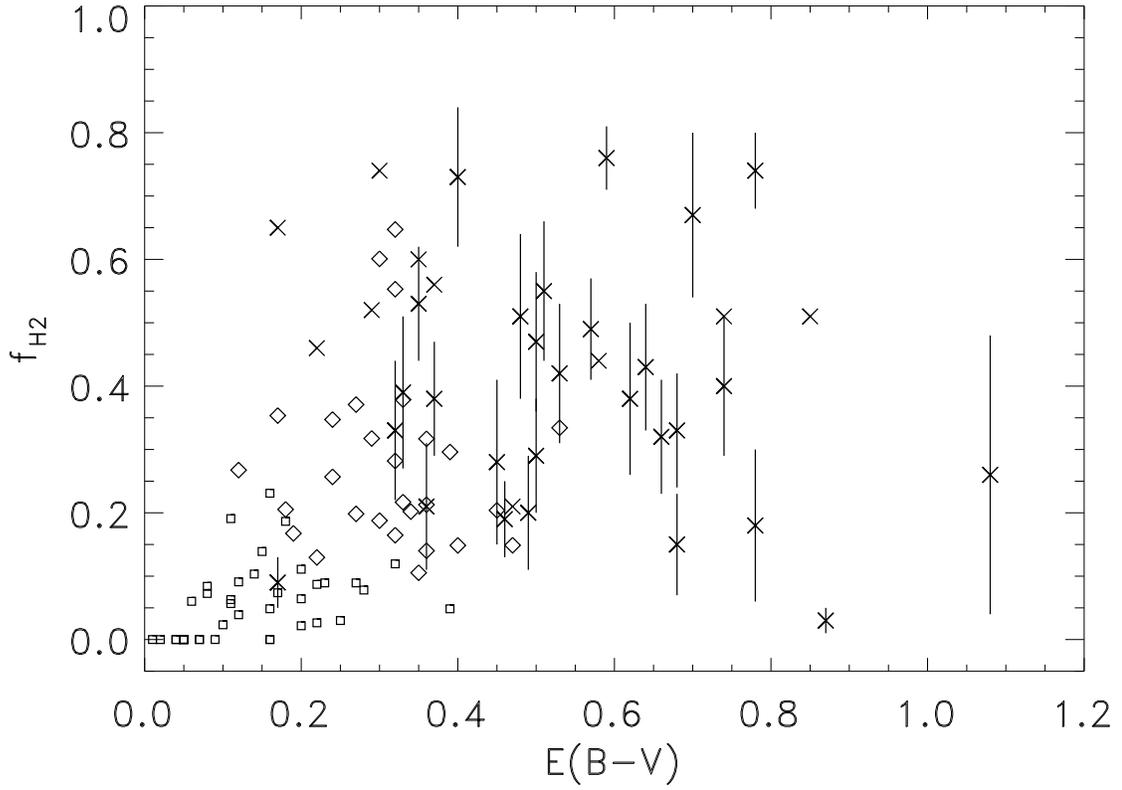}
\caption{Molecular fraction versus color excess.  Crosses: {\it
FUSE}; diamonds: {\it Copernicus} points with
$N$(H$_2$) $>$ 10$^{20}$ cm$^{-2}$; squares: {\it Copernicus}
points with $N$(H$_2$) $<$ 10$^{20}$ cm$^{-2}$.  Typical $f_{\rm
H2}$ errors for {\it Copernicus} data points ($\sim$30\%) are
slightly larger than for the {\it FUSE} data points.  Error
bars are not given for {\it FUSE} values derived without direct
measurements of $N$(H I).}
\end{figure}

\clearpage
\begin{figure}
\plotone{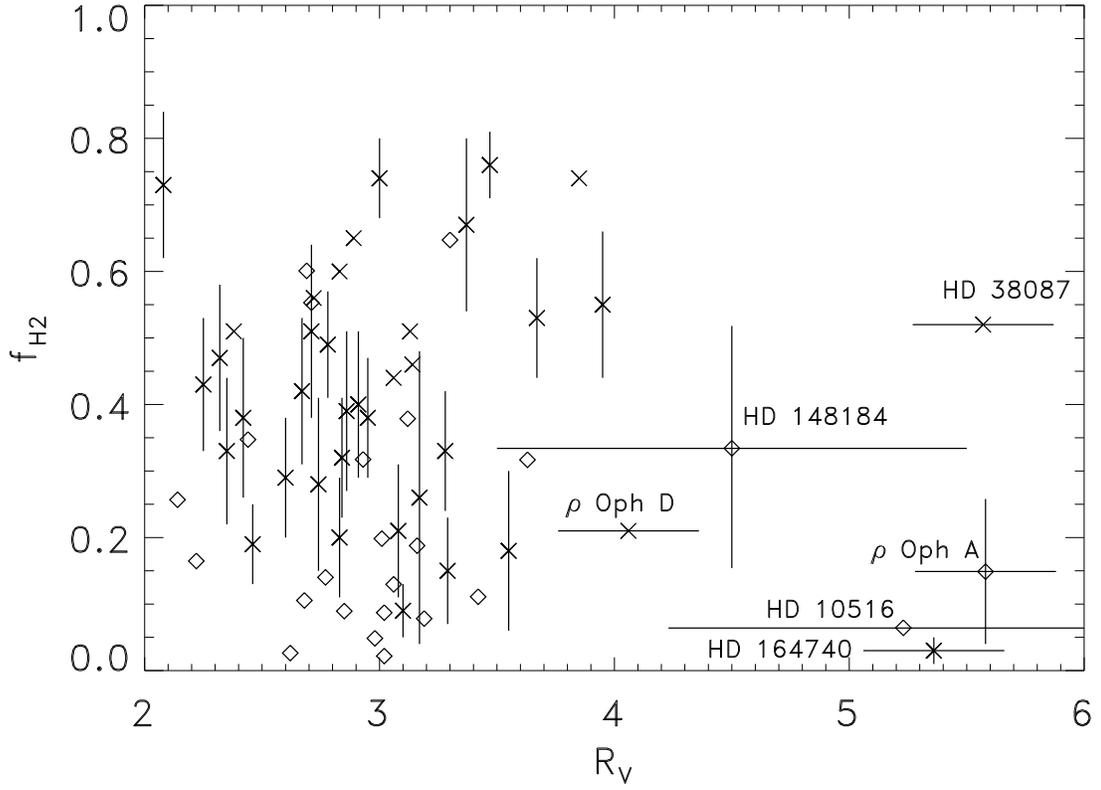}
\caption{Molecular fraction versus total-to-selective extinction
ratio.  Symbols and comments as in Figure 5.  To make the plot
easier to read, error bars for $R_V$ are only given for the high
$R_V$ sample discussed in \S\ 4.3.  We also give error bars for
$f_{\rm H2}$ for the two {\it Copernicus} targets discussed in
\S\ 4.3.}
\end{figure}

\clearpage
\begin{figure}
\plotone{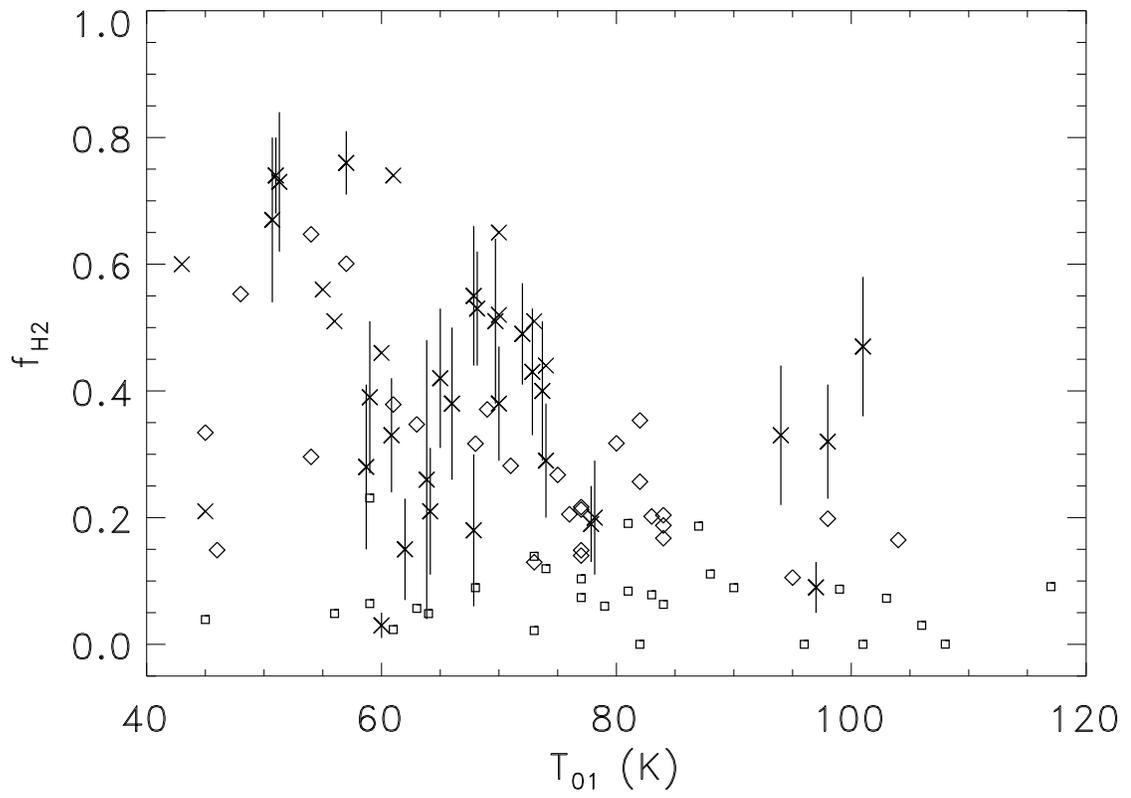}
\caption{Molecular fraction versus rotational temperature.  Symbols
and comments as in Figure 5}
\end{figure}

\end{document}